# The effect of heavy tars (toluene and naphthalene) on the electrochemical performance of an anode-supported SOFC running on bio-syngas

Davide Papurello*, Andrea Lanzini, Pierluigi Leone, Massimo Santarelli

[a] Department of Energy (DENERG), Politecnico di Torino, Corso Duca degli Abruzzi, 24, 10129, Turin, Italy.

**Abstract**

The effect of heavy tar compounds on the performance of a Ni-YSZ anode supported solid oxide fuel cell was investigated. Both toluene and naphthalene were chosen as model compounds and tested separately with a simulated bio-syngas. Notably, the effect of naphthalene is almost negligible with pure $H_2$ feed to the SOFC, whereas a severe degradation is observed when using a bio-syngas with an $H_2$:CO = 1. The tar compound showed to have a remarkable effect on the inhibition of the WGS shift-reaction, possibly also on the CO direct electro-oxidation at the three-phase-boundary. An interaction through adsorption of naphthalene on nickel catalytic and electrocatalytic active sites is a plausible explanation for observed degradation and strong performance loss. Different sites seem to be involved for $H_2$ and CO electro-oxidation and also with regard to catalytic water gas shift reaction. Finally, heavy tars (C≥10) must be regarded as a poison more than a fuel for SOFC applications, contrarily to lighter compounds such benzene or toluene that can directly reformed within the anode electrode. The presence of naphthalene strongly increases the risk of anode re-oxidation in a syngas stream as CO conversion to $H_2$ is inhibited and also $CH_4$ conversion is blocked.

*Corresponding author. Tel.:+393402351692. *Email address*: davide.papurello@polito.it

**Nomenclature:**

ASC, anode supported cell,

ASR, area specific resistance,

$C_{10}H_8$, naphthalene,

$C_2H_2$, acetylene,

$C_2H_4$, ethylene,

$C_7H_8$, toluene,

ECN, energy research centre of the netherlands,

EIS, electrochemical impedance spectroscopy,

ESC, electrolyte supported cell,

FU, fuel utilization,

LHV, low heating value,

Ni-GDC, nickel mixed with the ceramic material GDC,

Ni-YSZ, nickel mixed with the ceramic material YSZ,

ppm(v), parts per million by volume,

RWGS, reverse water gas shift reaction,

SEM/EDS, scanning electron microscope/energy dispersive x-ray spectrometry,

SOFC, solid oxide fuel cell,

TPB, three phase boundary,

WGS, water gas shift reaction.

**Introduction**

Solid oxide fuel cells (SOFCs) coupled with biomass source offer the potential of highly efficient combined heat and power generation [1]. The theoretical energy performance expected from integrated biomass gasifier and SOFC systems has been investigated by several authors [2–6] and efficiency as high as 50% (biomass-to-electricity, LHV basis) can be reached. However, the direct use of biosyngas may degrade the performance of SOFCs as it contains minor species, including particulates, hydrogen sulfide, chlorides, alkali compounds, and tars. These species can deactivate or damage SOFC anodes [4,7–10] when contained in the feeding stream. Among these trace species, tars have been identified as a major concern in developing gasifier–SOFC power systems as they can potentially deactivate the anode catalysts and degrade fuel cells performance mainly through carbon deposition [11]. Tar removal is still a critical issue in the effort toward integrated biomass-fed gasifier fuel cell systems [12]. The amount of tars in the syngas depends much on the gasifier configuration (up-draft, down-draft, fluidized circulating bed) as well as operating conditions (gasifier agent, temperature and pressure). A study of Phuphuakrat et al., (2010) showed for a downdraft dry sewage sludge gasifier, a syngas with the following tar content: essentially benzene, toluene, xylene, styrene, phenol and naphthalene [13]. Furthermore, when seeking for a hot integration of the gasifier with the SOFC generator, tar removal at high temperature becomes generally costly as a dedicated catalytic tar cracker reformer is mandatory. There is still the risk of trace amount of heavy hydrocarbons that passes through the catalytic bed to achieve the SOFC anode. In Aravind et al., (2008) was studied the impact of naphthalene (110 ppm(v)) on the performance of electrolyte supported cells with a graded Ni-GDC anode [7]. The cell showed feasible performance with that concentration. However, only a $H_2$-$H_2O$ anode feeding mixture was tested and no results were provided for CO-, $CH_4$- containing fuel. Recently, Liu et al., (2013) studied the effect of toluene on the performance of Ni-GDC anodes in electrolyte-supported type cells [14]. Toluene was well tolerated during the experiments up to 20 g/Nm$^3$ (corresponding to around 5,000 ppm(v) of contaminant in the anode syngas feed). The performance of Ni-GDC composite anodes is generally recognized to be more resistant to carbon deposition than Ni-YSZ, as also shown in the study of Lorente et al., (2013), where the two materials were tested with real tar fractions in a micro-reactor configuration [15]. The short-term operation (~7 hrs) on a high tar load (>10 g/Nm$^3$) has proven that tars did not cause immediate problems to Ni-GDC anode operation [7]. Indeed, no performance losses were observed and no carbon or other product gas trace constituents contamination of the anodes was found when the SOFC membranes were examined with SEM/EDS after the tests. To avoid carbon deposition it seems to be a synergic effect of having a thin Ni-GDC anode support on an electrolyte supports during the electrochemical operation. The Ni-phase is mostly involved (active) toward electrochemical reactions and it is not deactivated by carbon forming reactions or other surface adsorption phenomena. However, limited studies were conducted so far on Ni-YSZ anode-supported cells that still represent the dominant design for commercial systems, especially because lower temperature operation is feasible (650-750 °C) compared to ESC. The presented experimental results of stable SOFC operation on real heavily tar-laden biomass derived product gas are promising when compared with the findings reported from tests with synthetically introduced tar at comparable loads on similar anode material, where the introduction of different tar levels, i.e. naphthalene (0.29 – 6 g Nm$^3$), phenanthrene (1 g Nm$^3$), pyrene (0.2 g Nm$^3$) caused a rapid voltage drop within a few hours before leveling out at a lower voltage. Usually, tar model compounds chosen for SOFC performance testing are benzene or toluene [13–15]. Only a very limited number of studies, according to our knowledge, are devoted to heavier tars such phenols and naphthalene [2,16,17]. At ECN was tested ASC and ESC cells with various tars, including toluene (0.4 %vol.), naphthalene (525 ppm(v)), phenanthrene (126 ppm(v)) and pyrene (22 ppm(v)) at 750 °C, 0.3 A/cm², FU = 60% in a reformate gas mixture. Cell degradation was clearly observed during toluene feeding, and much steeper with naphthalene. However, a very high-tar containing anode feed was used, and safe limits of operation with such contaminants were not identified. In the same report, ECN also evaluated the performance of ESC cell with a Ni-GDC graded anode at 850 °C, 26%vol. $H_2O$, 0.16 A/cm², FU = 60%. In this case a higher tolerance to toluene was observed, still with a degradation trend. Naphthalene was inserted gradually this time, with concentrations of 50, 100, 250 and 525 ppm(v), respectively. Even with the lowest amount (50 ppm(v)), degradation was observed. In the same experiment, anode feeds containing phenanthrene (126 ppm(v)) and pyrene (22 ppm(v)) were also tested showing strong degradation. From such experiments, more refined with the ESC-type cell, the clearly emerging trend is that the higher the C amount (i.e., the heavier the tar), the higher is also the degradation. Carbon deposition was the recognized source of degradation, even though very fast deactivation occurs which may be due to also to other phenomena that we have tried to better understand with the present work. The effect of acetylene ($C_2H_2$) and $C_2H_4$ (ethylene) was also studied, resulting in no degradation when feeding up to 0.1 %vol. of $C_2H_2$ and 1.4 %vol. of $C_2H_4$. Again, lower C-compounds are easily converted as fuel in the SOFC. This is true also for toluene, however de-activation and related cell degradation start becoming evident. For heavier compounds (i.e., C $\geq$ 10), degradation becomes more marked and fuel conversion is strongly inhibited (also for lighter hydrocarbons present in the anode feed).

In this study, we have chosen toluene and naphthalene as tar model compounds to study the electrochemical performance of a commercial Ni-YSZ anode-supported SOFC running on simulated bio-syngas containing up to few ppm(v) of such hydrocarbons. Only few studies report the effect of such contaminants on the SOFC performance when running with such anode mixture. A real syngas fuel was reproduced and a full understanding of the interaction of naphthalene with a Ni-YSZ under relevant electrochemical operating conditions is still lacking. Our focus is limited to the impact of few ppm(v) (up to 50 ppm(v)) of naphthalene in the anode feed. We focus on commercial-type anode-supported Ni-YSZ cells, for which almost no literature studies are available for SOFCs fed by heavy tars in a syngas mixture during an electrochemical experiment.

**Material and methods**

The experiments are intended to simulate the behavior of wood gas from biomass gasification on a SOFC anode. In the test bench, synthetic wood gas consisting of $H_2$, $CO$, $CH_4$, $CO_2$ and $N_2$ that are mixed using five mass flow controllers (Bronkhorst, The Netherlands). A liquid mass-flow controller in connection with a controller mixer evaporator was used to feed and vaporize water, to be mixed with the other dry gaseous components. Naphthalene was co-fed with $H_2$ using a certified gas cylinder (Siad Spa, Italy) containing 50 ppm(v) of $C_{10}H_8$ in $H_2$. The cylinder pressure was limited to 28 bar(g) in order to avoid tar condensation and thus segregation from the gas matrix. Toluene was co-fed with $CH_4$ using a certified gas cylinder (Siad Spa, Italy) containing 200 ppm(v) of $C_7H_8$ in $CH_4$. The cylinder pressure was limited to 46 bar(g) in order to avoid tar condensation and thus segregation from the gas matrix. Such approach is useful when co-feeding, with anode mixture, low or ultra-low concentration of contaminants in an extremely accurate way. Nonetheless, conventional systems based on evaporator and carrier gas, where the contaminant is available as pure compound and a carrier gas is used to extract the vapor phase formed above the liquid (or solid) volume fraction, are the only option for higher concentrations. In this study, the interest was to find the threshold limit of heavy tar compounds affecting the SOFC performance. Moreover, in the real-configuration system, testing low concentrations could be also relevant as they could represent the case of a tar-cracker able to convert most of the tar in the producer gas, leaving however some hydrocarbons unconverted. Nonetheless, the main motivation of using low concentrations to investigate the underlying degradation phenomena occurring in the presence of heavy-tars in the anode feed stream. A commercial anode-supported cell (SOLIDpower Spa, Italy) was used for all experiments. The cells consist of a Ni-YSZ anode support of ~240 μm, an anode active layer, a YSZ dense electrolyte and a LSCF cathode. The cells consisted of a disk of 50 mm of anode diameter which were tested in oven configuration with seal-less ceramic housings. A detailed description of the experimental apparatus can be found elsewhere **Errore. L'origine riferimento** **non è stata trovata.**. The operating temperature was 740 – 750 °C: the cell temperature was measured through a thermocouple (TC) placed at the center of the cell on the anode side. In this way, the monitored TC signal is quite representative of variation occurring on the anode electrode due to chemical reactions (e.g., endothermic steam-reforming, slightly exothermic water-gas shift, etc.). The volumetric dry syngas composition used for all experiments is $H_2/CO/CH_4/CO_2/H_2O/N_2$ 19/17/2/8/7/47, yielding an $H_2/CO$ ratio quite close to 1. A (molar) steam-to-carbon of 0.4 (related to amount of carbon contained in C-fuels, i.e., CO and $CH_4$) was set to prevent carbon deposition while avoiding an excessive amount of steam that could lead to enhanced Ni degradation. On another test bench, the catalytic conversion study of a gas mixture with tars was carried out with a concentration of $C_{10}H_8$ on nickel anode half-cell. The reactor was placed into a furnace whose operative temperature was leaded by a controller driven (Horst Gmbh, Germany) by a K–type thermocouple placed in a thermo well cantered in the catalyst bed. Two additional thermocouples (K–type) were located at the inlet and outlet of the catalytic bed to measure temperature of inlet and outlet gas flows (Tersid, Milano Italy). The heating apparatus was calibrated in such a way that all temperature measuring points had the same constant temperature. An alumina fixed-bed reactor ($\Phi_{int.}$=14 mm; $h_{bed}$=180 mm) with a previously reduced anode + electrolyte cell was heated up to 700 °C in a ceramic electric furnace. The reactor was fed by two different inlet streams:

- $H_2/CO_2$ (50/50 %vol.) mixture with 25 ppm(v) of $C_{10}H_8$,
- The syngas mixture with a steam to carbon ratio fixed at 0.4 with 10 ppm(v) of pollutant (concentration selected as the maximum value possible depending on the $H_2$ molar concentration adopted).

According to Guerra et al., (2014), we selected the fuel flow and the nickel anode amount: the nickel anode half-cell reduced was about 0.55 g [18]. An HPR-20 mass spectrometer (Hiden ltd, UK) has been adopted to monitor the furnace outlet gas concentration.

## Results

**Naphthalene effect on SOFC performance**

The effect of Naphthalene was first check against an anode feed containing $H_2$-fuel only, humidified at 6.1%. The fuel cell performance was not affected by the presence of a varying amount of $C_{10}H_8$ up to 50 ppm(v) under an operating temperature of 750 °C and a current density of 0.33 A/cm$^2$ (fig. 1). In fig.2 the EIS were measured during the galvanostatic experiment shown in fig. 1 (portion of the graph on the left, where $H_2$ fuel is used). Essentially the impedance response was observed with/without the tar. Actually, a slightly shift on the left (meaning lower overall polarization) was observed when feeding naphthalene as a result of an activation process still occurring. Adding at the hydrogen stream, naphthalene the test improves due to the more fuel adopted that is reformed on the cell. Considering the intercept with the real axis a slight improvement is recorded for the $H_2$ + $C_{10}H_8$ case. However, the shape of circles observed is essentially the same, underlying that no interaction of tar with electrochemical performance were observed with $H_2$ fuel only at such low tar concentrations in the feeding fuel. In fact, the ASR value is 0.52 $\Omega$cm$^2$ with pure $H_2$ and 0.49 $\Omega$cm$^2$ adding 25 ppm(v) of $C_{10}H_8$. This is also in agreement to what found in **Errore. L'origine riferimento non è stata trovata.**. The response was completely different when feeding the cell with the syngas mixture, humidified at 14%, with a varying amount of naphthalene up to 10 ppm(v). As the naphthalene concentration gets higher, a sudden voltage drop is observed followed by stable voltage operation up to ~5 ppm(v). For concentrations above 5 ppm(v), both initial voltage drop and subsequent performance degradation becomes steep. At 10 ppm(v), the voltage drop is such to reverse the potential of the fuel cell indicating a strong re-oxidation occurring on the Ni-anode. In fig. 3 the portion of experiment when naphthalene is added to the synthetic syngas stream is enlarged, showing well how the cell temperature results reduced by almost 20 °C while switching from tar-free syngas to a maximum concentration of $C_{10}H_8$ of about 10 ppm(v). Naphthalene is recognized to slow/inhibit steam-reforming kinetics of methane fuel.

However it is not clear its effect on water-gas shift reactions considering that a potential is also applied to the electrode. Considering that the syngas contains both CH₄ (2%vol.) and CO (15%vol.), the involved chemical reactions within the Ni-YSZ anode electrode are:

$$CH_4 + H_2O \leftrightharpoons 3H_2 + CO, \ \Delta H^0 = 206 \ kJ/mol \quad \textit{steam-reforming (SR)} \ (1)$$

$$CO + H_2O \leftrightharpoons H_2 + CO_2, \Delta H^0 = -41 \ kJ/mol \quad \textit{water-gas shift (WGS)} \quad (2)$$

The temperature drop in fig. 3 when inserting $C_{10}H_8$ in the anode syngas stream is consistent with a strong inhibition of the WGS reactions (Eq. 2) rather SR suppression, that would have caused an opposite temperature behavior. The regular oscillation observed in the temperature profile is noise from the electronic instrumentation used for data acquisition.

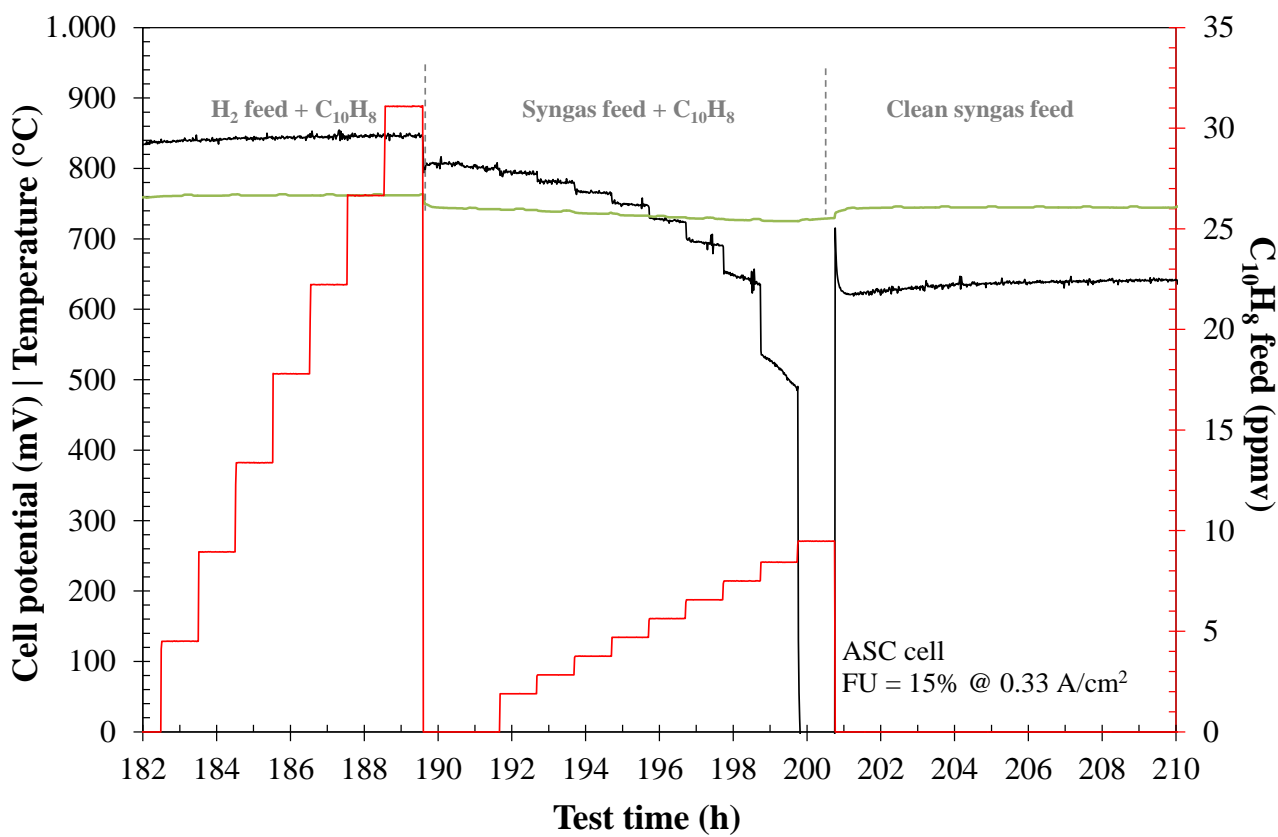

**Figure 1 – The effect of naphthalene ($C_{10}H_8$) on the ASC performance when feeding an $H_2$-$H_2O$ only fuel and bio-syngas.**

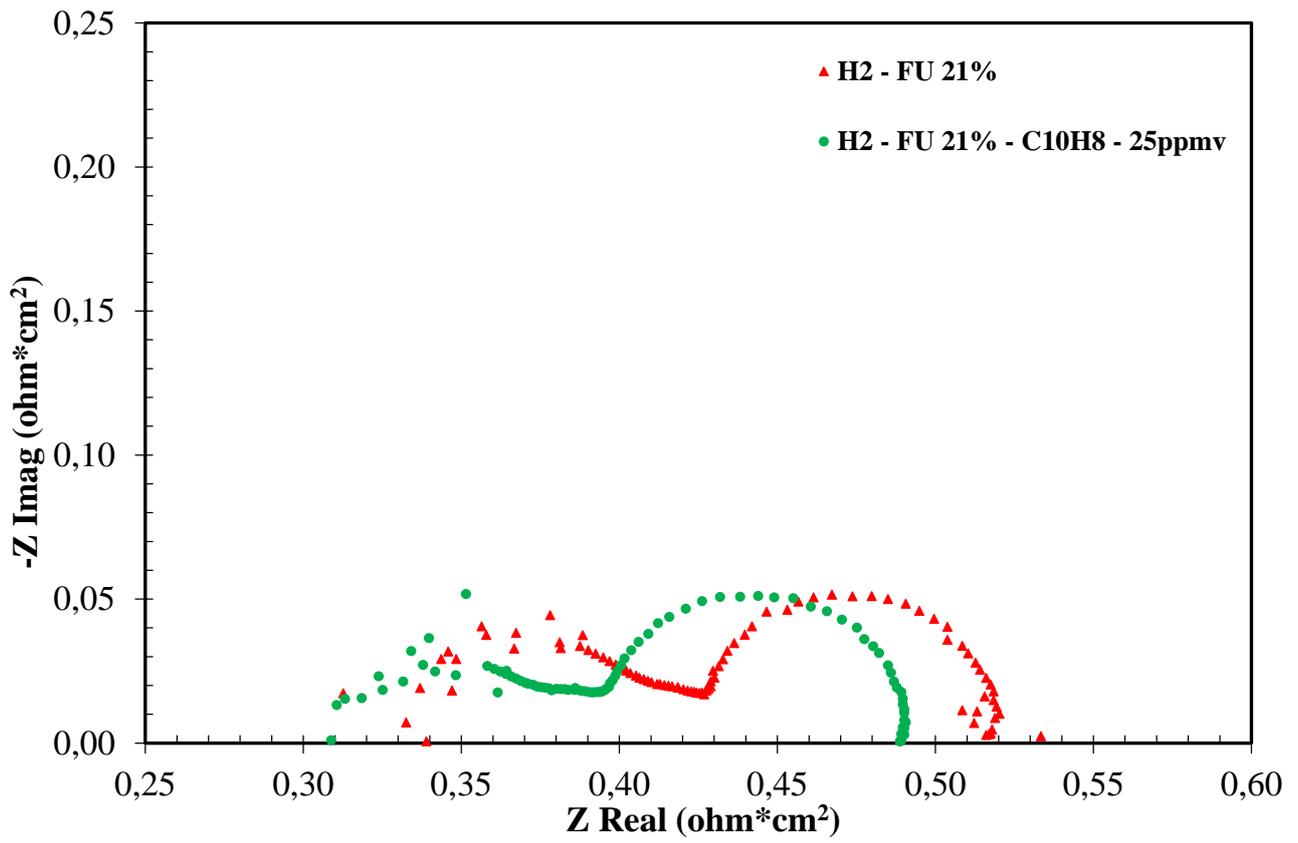

**Figure 2 – EIS with clean $H_2$ fuel and $H_2$ + 25 ppm(v) of naphthalene (current density 0.33 A/cm$^2$, temperature 750 °C).**

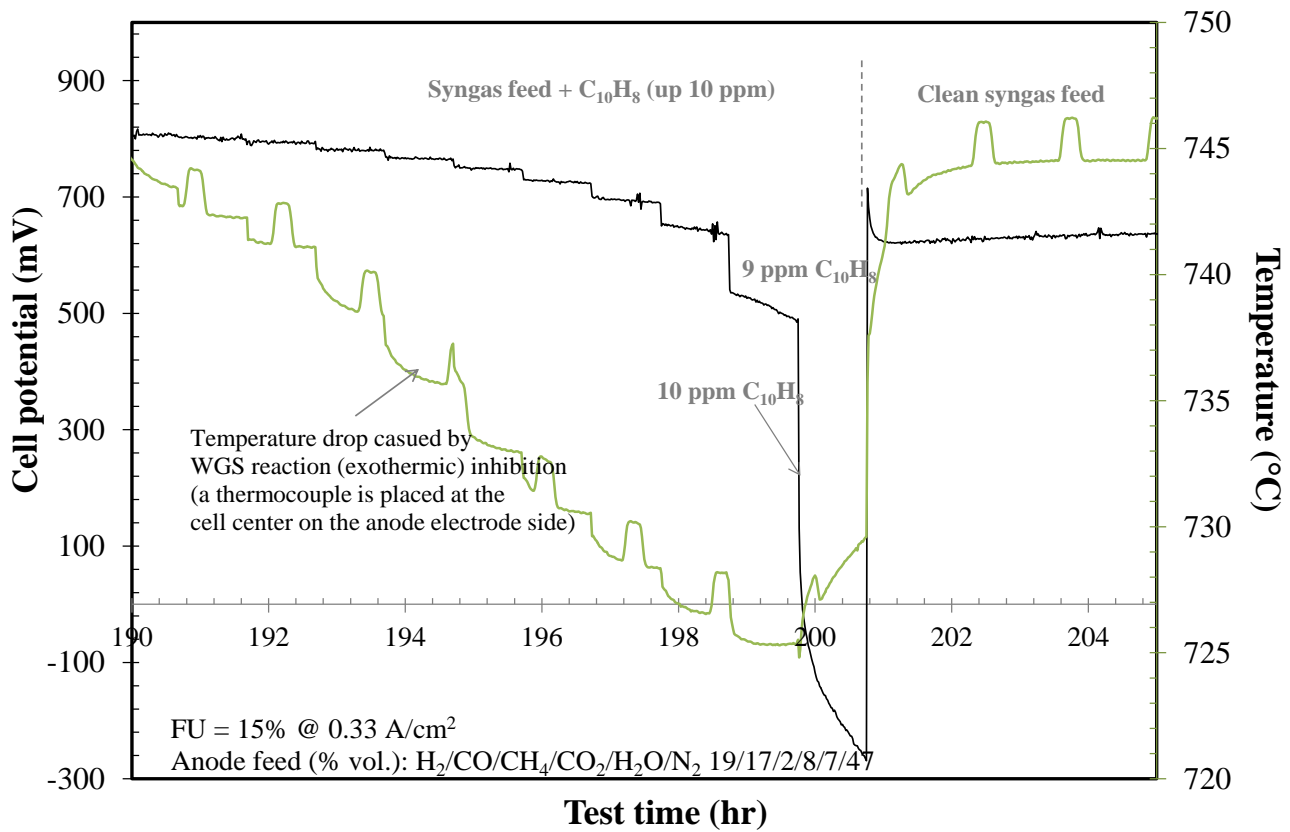

**Figure 3** – The effect of naphthalene ($C_{10}H_8$) on ASC performance when feeding syngas. The temperature behavior is also plotted.

In order to further understand the impact of naphthalene on the WGS reaction, and thus to analyse what stated before, polarization experiments were run using anode fuel compositions without methane (fig. 4). In this way, the only relevant chemical reaction occurring on the Ni-anode is the WGS reaction (or its reverse). Being the naphthalene concentration always around few ppm(v)s, the tar in the anode feed is always considered as a fuel impurity rather than a fuel able to provide a meaningful amount of electricity. However, the fate of naphthalene remains unclear and should be further investigated with gas analysis of the anode outlet. Polarization results show and confirm that the impact of heavy tar (naphthalene in our study) does not affect significantly the fuel cell performance under $H_2$ fuel only. Performance are only slightly reduced with naphthalene at higher current density, not necessarily meaning an effect of the tar. The performance is within the variability observed among repetition of the i-V trace under same operating and fuel conditions.

The performance with H$_2$/CO 70/30 %vol. in absence of any steam was instead producing quite different results depending on the presence of tar. In fact, naphthalene strongly reduced the cell performance. Most notably, already the OCV is reduced, when switching from a clean H$_2$/CO feed to one with tar showing that, even at open circuit WGS is inhibited thus reducing the H$_2$ partial pressure at anode electrode. The performance with CH$_4$ and H$_2$ plus steam with and without the pollutant impact of naphthalene are reported in figure 4. A strong degradation is reported especially increasing the current density required, the reforming of methane is supposed to be limited with a decreasing on the performance around 32%.

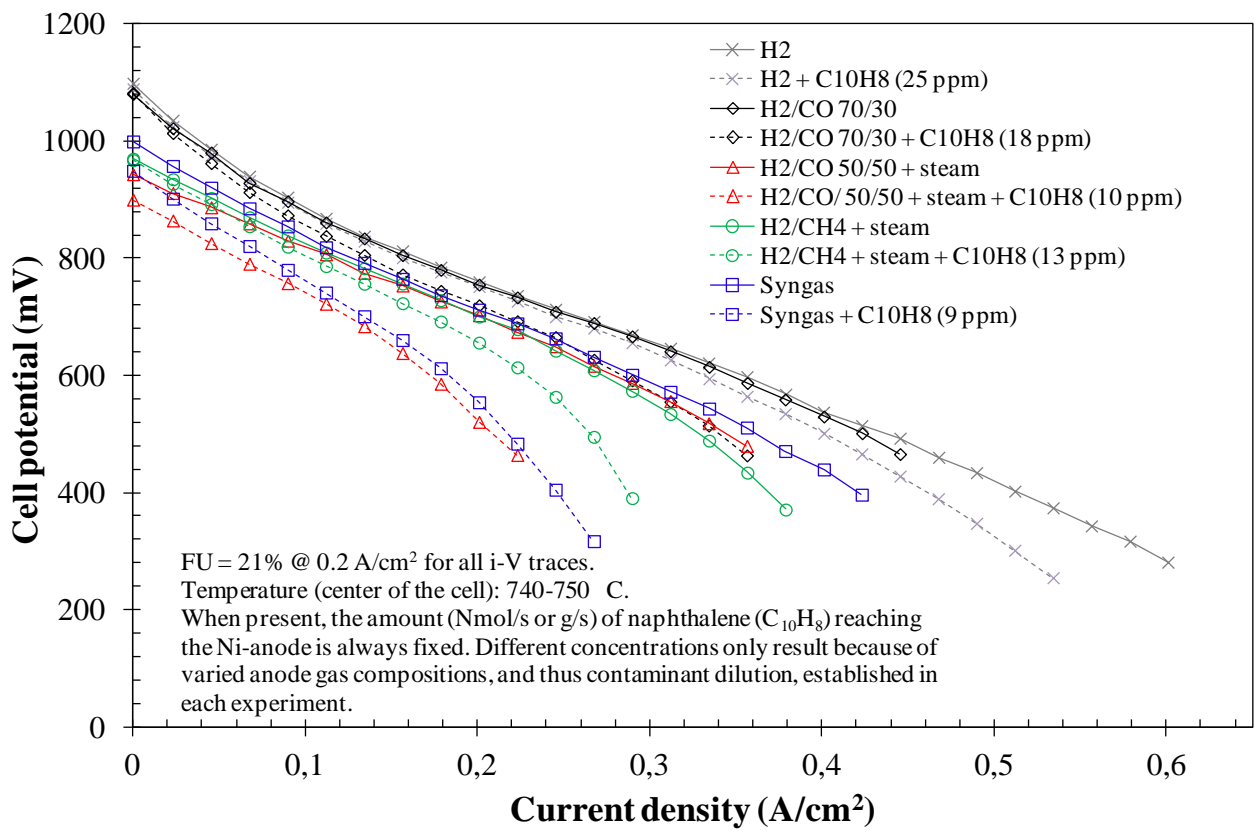

**Figure 4 – Polarization curves with/without the presence of naphthalene under different anode fuel compositions (pure H$_2$, H$_2$/CO 70/30, H$_2$/CO 50/50 and H$_2$/CH$_4$).**

**Toluene effect on SOFC performance**

Due to results achieved in the previous section toluene was tested only with syngas mixture in order to investigate the tar effect on SOFC performance. As pointed out in figures 5 and 6, when feeding the syngas mixture with two different amount of toluene, 3.8 and 24.2 ppm(v) the response was completely different. When 3.8 ppm(v) of toluene was added to the syngas, the EIS behavior and cell voltage value were slightly higher than the values recorded in clean syngas. In figure 5 (galvanostatic mode) it is shown that for almost 100 hrs the cell performance has been improved, as the toluene acts as a fuel decreasing the FU. The cell voltage passes from almost 780 mV with clean syngas up to 830 mV for 60 h and 820 mV stable, for 80 h with syngas adding 3.8 ppm(v) of toluene. This voltage decreasing is related to an unstable behavior of the cell with a low toluene concentration. Adding 24.2 ppm(v) of toluene in the syngas mixture, the cell performance shows a strong deterioration. The cell voltage remains quite stable to 800 mV for 60 h, after that a strong decrease is reported. A decrease of almost 0.57 mV/h is depicted. This behavior is in accordance with the behavior recorded in case of naphthalene, demonstrating how syngas + tar (relative low concentration) can be detrimental for SOFC applications.

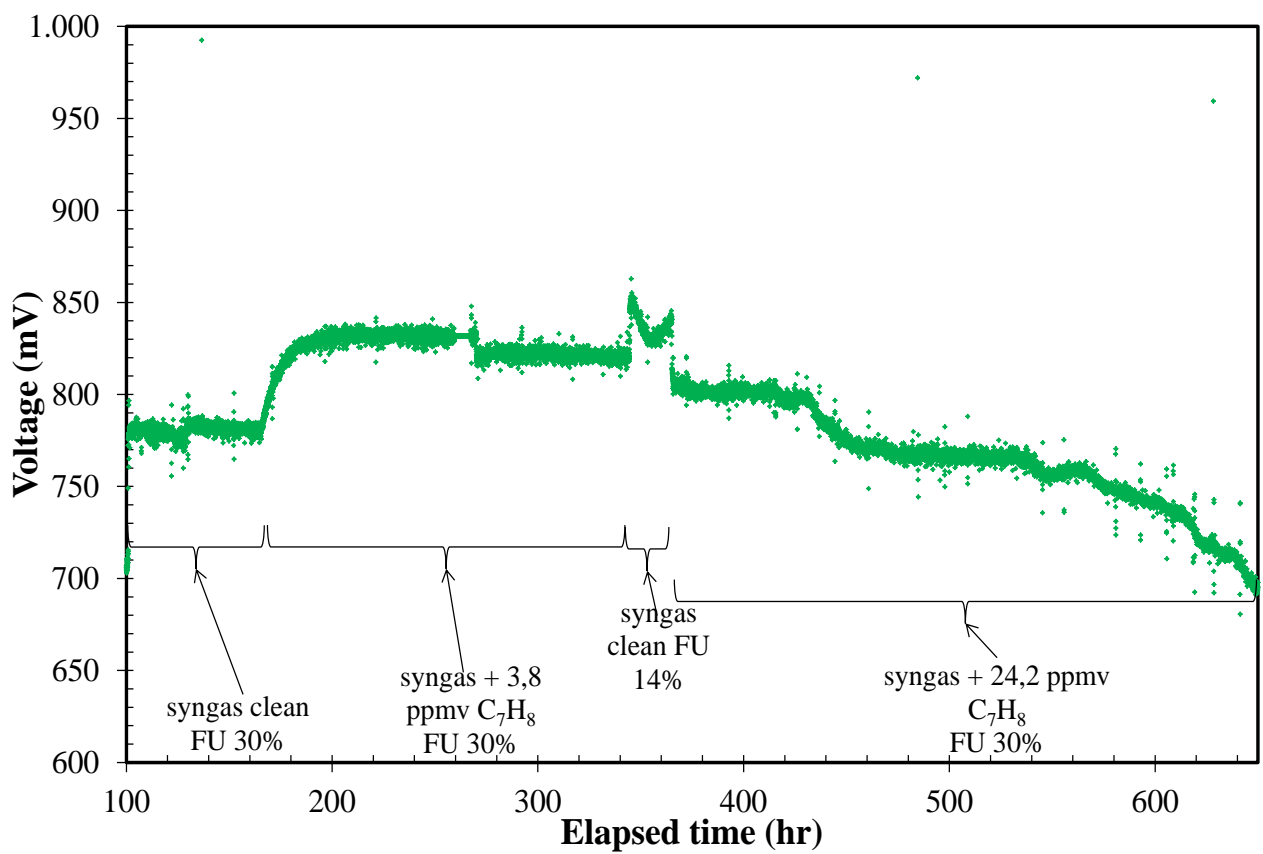

**Figure 5 – Cell voltage curves (galvanostatic mode) with/without the presence of toluene under syngas fuel composition.**

A reversible behavior for toluene was reported for concentrations below 10 ppm(v). In fact EIS curves for clean syngas and syngas + 4 ppm(v) of $C_7H_8$ show similar behavior (figure 6) while with the concentration around 24 ppm(v), a strong performance decrease is reported. Figure 6 describes the EIS values for pure $H_2$, syngas at steam to carbon of 0.4 and 0.35. Moreover it describes the EIS behavior of syngas plus 3.8 ppm(v) and 24.2 ppm(v) of toluene. Comparing the behavior of the cell fed by $H_2$ with clean syngas at S/C 0.35 and 0.4 a strong influence, especially on the polarization resistance are reported. The ohmic voltage for H2 is 0.18 $\Omega cm^2$ comparable with 0.22 $\Omega cm^2$ for the syngas mixture. All the syngas mixtures with and without toluene show the same ohmic value. The ASR discrepancy for syngas (S/C 0.35) and pure hydrogen is around 38%. Increasing the steam to carbon value the ASR increases of almost 27%. This is mainly due to the more dilution effect of water decreasing the electrochemical fuel available. Adding to the syngas mixture 3.8 ppm(v) of toluene, the AR value remains almost unchanged with the clean fuel. This demonstrates how the cell behavior is not affected by toluene at such concentrations. Increasing the toluene concentration the ASR increases due to problems related to the syngas direct internal reforming on the cell, hampered by the difficulty of reforming such tar compound with a syngas mixture. A similar behavior was recorded when it was tested naphthalene (10 ppm(v)) with a syngas mixture.

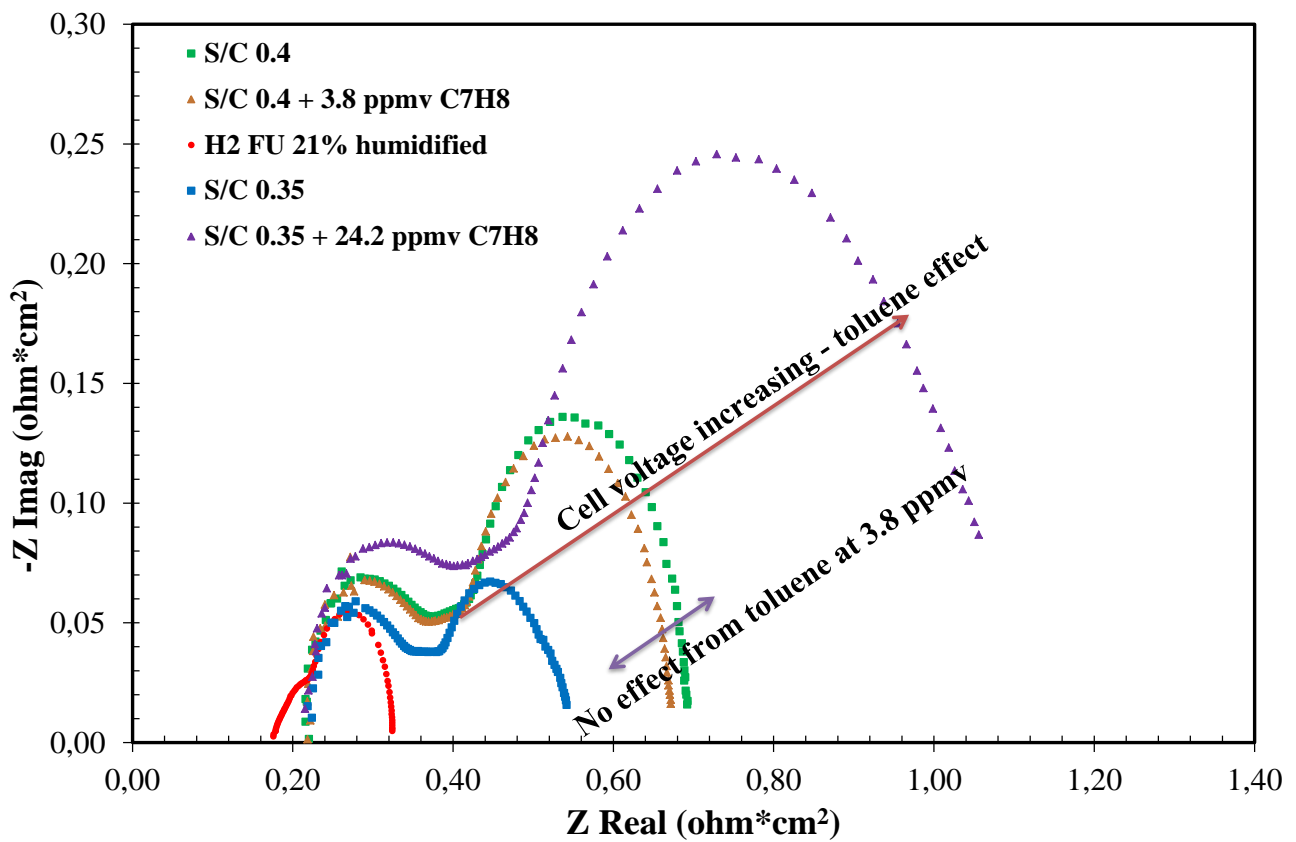

**Figure 6 – EIS with clean syngas and syngas + 3.8 and + 24 ppm(v) of toluene (current density 0.33 A/cm$^2$, temperature 750 °C).**

Considering the method investigated by Papurello et al., (2016), figure 7 shows the different contribution on the ASR value for the syngas with and without the toluene concentration [19].

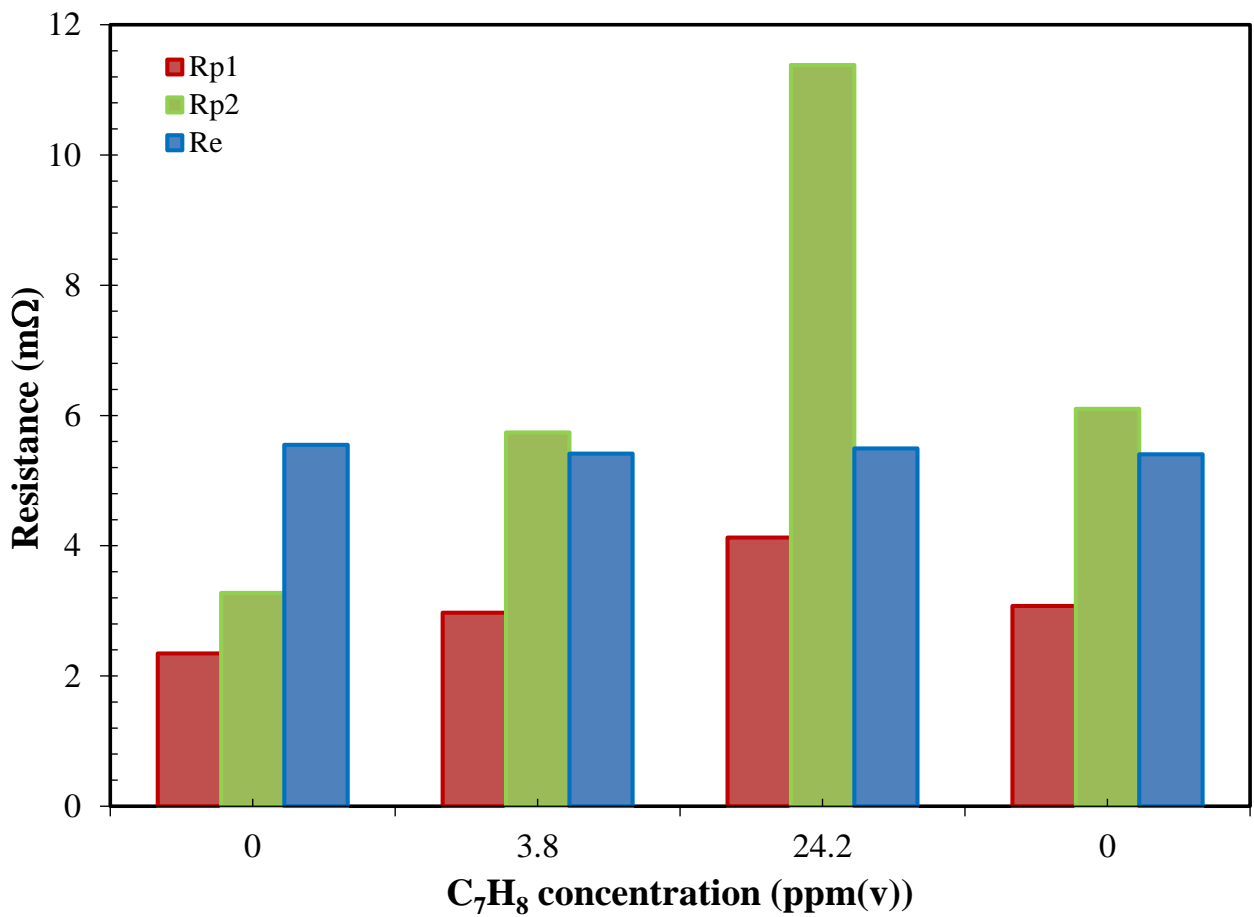

**Figure 7 – EIS contribution for clean syngas and syngas + 3.8 and + 24 ppm(v) of toluene (current density 0.33 A/cm$^2$, temperature 750 °C).**

The ohmic contribution, Re, remains stable for all the case studied. The most affected term is Rp2, increasing the toluene from 3.8 ppm(v) to 24.2 ppm(v). This value is related to the low frequency term that considers mass transport phenomena. Removing from the syngas mixture the ohmic contribution remains stable, while Rp2 change.

**Naphthalene catalytic study on Nickel anode compartment**

The anode nickel based fuel cell have been fed with $H_2/CO_2$ and syngas mixtures at 700 °C in a catalytic reactor (without electrochemical reactions). The goal of this section is to verify the difficulty to complete the reverse water gas shift reaction (RWGS) (eq. 3) using naphthalene as tar model in the gas mixture.

$$CO_2 + H_2 \leftrightharpoons H_2O + CO, \Delta H^0 = 41\ kJ/mol \quad reverse\ water\text{-}gas\ shift\ (RWGS) \quad (3)$$

Figure 7 shows the gas concentration at the reactor outlet in case of $H_2/CO_2$ 50/50 %vol. inlet mixture at 700 °C with and without naphthalene.

In accordance with results achieved on electrochemical tests, the catalytic study showed how 25 ppm(v) of $C_{10}H_8$ decrease the CO and water concentration content in the reactor outlet. This demonstrate how naphthalene strongly reduces the cell performance via the RWGS inhibition: reaction (3) consumes $H_2$ and $CO_2$ to produce additional CO at clean conditions, while considering the case with naphthalene (10 ppm(v)) the CO and $H_2O$ concentration decreases whereas increase the concentration of hydrogen.

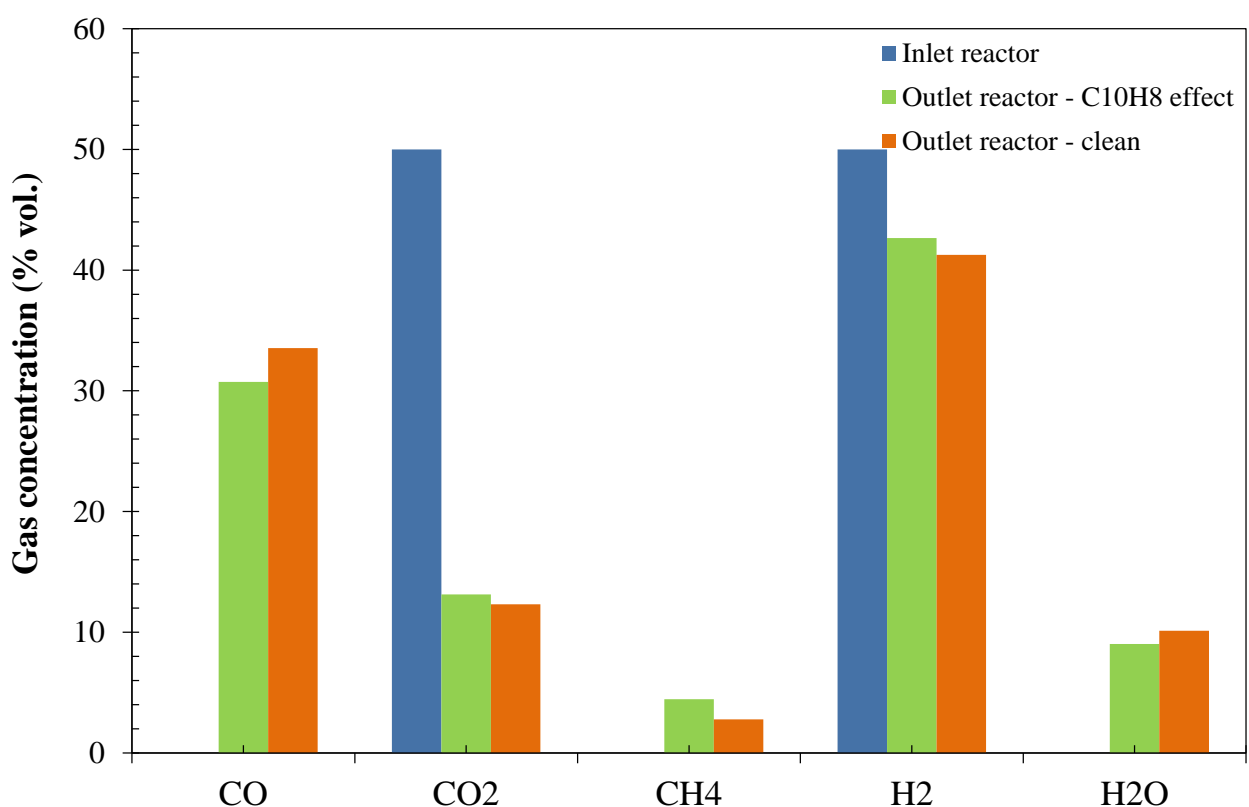

**Figure 8 – Gas outlet concentration for $H_2/CO_2$ mixture at 700 °C.**

The decreasing rate of CO and $H_2O$ is more pronounced in the syngas case where a direct internal reforming has to be performed on nickel active sites, see figure 8. These results are in accordance with those achieved during the electrochemical tests. The RWGS reaction is thus inhibited increasing the $H_2$ partial pressure at the anode electrode.

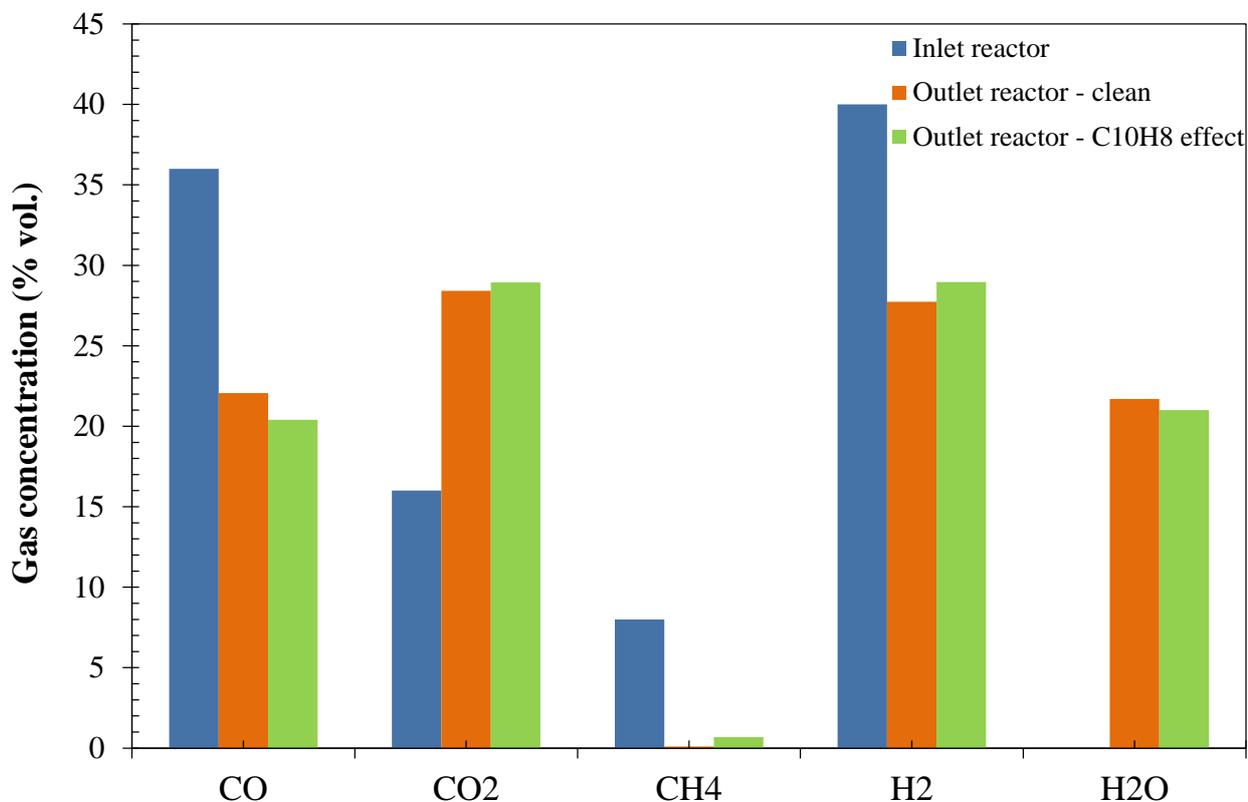

**Figure 9 – Gas outlet concentration for syngas mixture at 700 °C.**

**Conclusions**

Assuming the SOFC operating at a constant current density the fuel utilization depends very much on the reforming rate. Since the steam reforming and WGS reactions occurs on the Ni catalyst surface according, it is possible that adsorbed hydrocarbons in case of slow conversion reaction kinetics – as it is the case of Naphthalene – they also slow electrochemical reactions by occupying active sites of the TPB. The steps of the reforming reaction mechanism are: adsorption of methane on the Ni surface, decomposition, surface reaction and desorption of the products from the Ni surface. The kinetic inhibition is due to the many steps of the reforming reaction that can lead to a decreasing reactive surface caused by the adsorption on the Ni particle. The adsorption and desorption of naphthalene is obviously interfering with the reforming of methane and WGS (i.e.,

CO conversion to $H_2$) by decreasing the reactive surface, and increasing the actual FU experienced by the fuel cell. The latter fact could lead to anode re-oxidation that would be very critical as yielding a reduced overall SOFC lifetime.

**Acknowledgements**

The research leading to these results has received funding from the Italian Minister for Education (MIUR) under the national project PRIN-2009 "Analisi sperimentale ed energetico/strategica dell'utilizzo di syngas da carbone e biomassa per l'alimentazione di celle SOFC integrate con processo di separazione della $CO_2$" and part of the SOFCOM project which is carried out by Politecnico di Torino and other European partners (FCH-JU) (www.sofcom.eu).